\documentclass[twocolumn]{article}
\usepackage[T1]{fontenc}
\usepackage{geometry}
\usepackage{graphics}

\makeatletter

\newcommand{\LyX}{L\kern-.1667em\lower.25em\hbox{Y}\kern-.125emX\spacefactor1000}

\usepackage[T1]{fontenc}
\usepackage[latin1]{inputenc}
\usepackage{geometry}
\usepackage{graphics}


\begin{document}

\noindent\textbf{Non-Fermi liquid angle resolved photoemission lineshapes of
Li\( _{0.9} \)Mo\( _{6} \)O\( _{17} \)} \ \ A recent Letter by Xue \emph{et
al.~}\cite{Xue} reports observations from angle resolved photoemission spectroscopy
(ARPES) on quasi one-dimensional (q-1D) Li\( _{0.9} \)Mo\( _{6} \)O\( _{17} \)
that above its \( T_{X} \) \( \approx  \)\( 24 \)K transition a peak dispersing
to define its Fermi surface (FS) develops Fermi energy (\( E_{F} \)) weight
requiring a Fermi liquid (FL) description. This finding contradicts our report
\cite{Denlinger} of a non-FL lineshape in this material. The reasoning in \cite{Xue}
that this new finding was enabled by improved angle resolution is flawed. Rather,
we point out here that the data of \cite{Xue} have fundamental differences
from other ARPES results and also band theory. Therefore, until the origin of
these differences are learned, the claims of \cite{Xue} must be held in abeyance.
These claims also include the report of an 80meV gap below \( T_{X} \), which
contradicts the zero gap found in optical spectroscopy \cite{Degiorgi} and
magnetic susceptibility and exceeds by more than two orders of magnitude the
value (0.3meV) implied by a gap-model interpretation of the resistivity rise
below \( T_{X} \) \cite{Greenblatt}\@. 

Improved angle resolution is not relevant for the claimed FL lineshape. Because
the \textbf{k}=\textbf{k\( _{F} \)} lineshape for both the FL and the Tomonaga-Luttinger
(TL) model (with \( \alpha <1 \)) are singular at \( E_{F} \), the \( E_{F} \)
weight for both increases steadily as the angle resolution improves. Indeed,
it is well known that one must angle-integrate (\textbf{k}-sum) to test for
the surprising difference between the Fermi edge of the FL and the \( E_{F} \)
power law onset of the TL model. Xue \emph{et al}.~sum ARPES data along the
q-1D \( \Gamma  \)--Y direction over \( \Delta \mathbf{k} \)=0.2\AA\( ^{-1} \)
and report a Fermi edge, whereas we always find only a power law onset at \( E_{F} \)
in angle summed spectra, including our new high resolution spectra shown below.
This difference arises because the data are fundamentally different.

\begin{figure}
{\centering \includegraphics{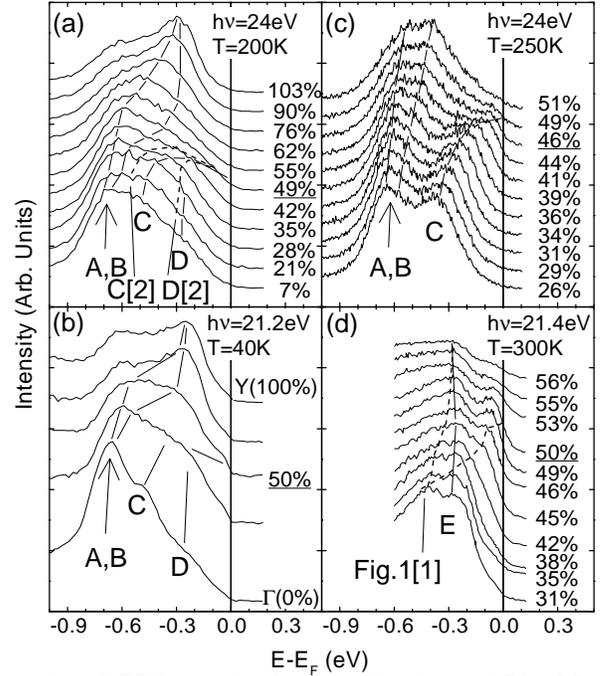} \par}
\vspace{-0.5cm}

\caption{ARPES data in the metallic phase of Li\protect\( _{0.9}\protect \)Mo\protect\( _{6}\protect \)O\protect\( _{17}\protect \)
along \protect\( \Gamma \protect \)\protect\( \rightarrow \protect \)Y\@.
Spectrum label, underlined at \textbf{k=k}\protect\( _{F}\protect \), is percentage
of the distance \protect\( \Gamma \protect \)Y\@. Energy and angle resolutions
are (a) 100 meV, \protect\( \pm 1^{o}\protect \), (b) 15 meV, \protect\( \pm 1^{o}\protect \),
{[}\ref{Grioni}{]}(c) 35 meV, \protect\( \leq \pm 0.25^{o}\protect \) and
(d) 33 meV, \protect\( \pm 0.1^{o}\protect \) {[}\ref{Xue}{]}. Long dashed
lines are guides to the eye for dispersions, and short dashed lines in (a) and
(d) are the C,D dispersions given by the data of {[}\ref{Denlinger}{]} and
the two dispersions reported in {[}\ref{Xue}{]}, respectively. \label{Figure}}
\vspace{-0.5cm}
\end{figure}

Fig.~\ref{Figure} shows various \( \Gamma  \)--Y data sets. The \textbf{k\( _{F} \)}
values show small variations among the data sets and so we place importance
on the \textbf{k} value relative to \textbf{k\( _{F} \)}\@. Panel (a) shows
previously unpublished data taken by us at photon energy h\( \nu  \)=24 eV
on literally the same cleaved surface as for the data of \cite{Denlinger}.
Panel (b) shows earlier data with h\( \nu  \)=21.2 eV by Grioni \emph{et al}.
\cite{Grioni} The data sets for the two h\( \nu  \) values are consistent,
both showing bands A through D in good basic agreement with band theory, as
labeled. Only bands C,D cross \( E_{F} \), becoming degenerate before the crossing.
In addition, there is an unexplained tendency, seen also in other Mo bronzes,
for non-dispersive weight to cling to the bottom of the band (\( \approx  \)\( -0.6 \)eV).
Panels (a) and (b) establish the consistency of the data of \cite{Denlinger}
and \cite{Grioni}. The special \textbf{k}-path parallel to \( \Gamma  \)--Y
in \cite{Denlinger} was chosen because both bands C,D are especially strong
all the way to \( E_{F} \)\@. The bands C,D in (a)--(c) along \( \Gamma  \)--Y
are in basic agreement with those in \cite{Denlinger} (short dashes in (a)),
and, importantly, show non-FL lineshapes as do the data of \cite{Denlinger}.
A difference occuring for the filled band B along the two paths is implied by
band theory but is not important here.

Panel (c) shows h\( \nu  \)=24 eV data taken at the Wisconsin Synchrotron Radiation
Center PGM beamline with an SES-200 Scienta analyzer over a narrow \textbf{k}-range
near \textbf{\( \mathbf{k}_{F} \)} where D is already too weak to see. These
data have angle and energy resolution comparable to that of Xue \emph{et al.}~and
fully agree with the data of (a) and (b) apart from generally increased sharpness,
and increased \( E_{F} \) weight for \textbf{\( \mathbf{k}_{F} \)}. It is
then meaningful to compare the data of (c) directly to the h\( \nu  \)=21.4
eV data of Xue \emph{et al.}~\cite{Xue}, shown in panel (d). Relative to \textbf{\( \mathbf{k}_{F} \)},
the two \textbf{k}-ranges are nearly the same. It is obvious by inspection that
the two data sets are globally different, for example by the absence in (d)
of peaks A,B and by the presence in (d) of a non-dispersing feature E which
interferes with the presumed C,D lineshapes, and has no counterpart in the other
data or in band theory. This work was supported by the U.S. DoE at U. Mich.\
and Iowa State, and by the U.S. NSF at U. Mich. and U. Wisc. 

G.-H.\ Gweon\( ^{1} \), J.D.\ Denlinger\( ^{1} \), J.W.\ Allen\( ^{1} \),
C.G.\ Olson\( ^{2} \), H.\ H\"ochst\( ^{3} \), J.\ Marcus\( ^{4} \) and C.\
Schlenker\( ^{4} \)

\emph{\( ^{1} \)Randall Lab.~of Physics, U.~of Michigan}, \emph{Ann Arbor,
MI 48109}. \emph{\( ^{2} \)Ames Lab., Iowa State U., Ames, Iowa 50011. \( ^{3} \)U.~of
Wisconsin}, \emph{SRC, Stoughton, WI 53589. \( ^{4} \)LEPES, CNRS, BP 166,
F-38042 Grenoble Cedex 9, France.}

\vspace{-1cm}

\end{document}